\documentstyle[12pt]{article}

\renewcommand{\in}{\raise -3pt\hbox{\scriptsize in}}
\newcommand{\out}{\raise -3pt\hbox{\scriptsize out}}

\newcommand{\PP}{P}        

\newcommand{\dist}{\Phi}

\newcommand{\eqcm}{\: ,}   
\newcommand{\eqpt}{\: .}

\begin{document}

\begin{flushright}
CPHT--S 045.0400
\end{flushright}

\begin{center}
\vskip 3.5\baselineskip
{\bf \Large Single spin asymmetries in
$\gamma^* \gamma \to \pi \pi \pi $ at large $Q^2$.}
Draft - June 2, 2000
\vskip 2.5\baselineskip
B. Pire$^{1}$ and O.V. Teryaev$^{1,2}$
\vskip \baselineskip
1. CPhT\,\footnote{Unit{\'e} mixte C7644 du CNRS}, \'Ecole
   Polytechnique, 91128 Palaiseau, France \\
2. Bogoliubov Laboratory of Theoretical Physics, JINR, \\
   141980 Dubna, Moscow region, Russia\,\footnote{Permanent address}
\vskip 3\baselineskip
{\bf Abstract} \\[0.5\baselineskip]
\end{center}

 The $\gamma^* \gamma \to \pi \pi \pi $ reaction is analysed in QCD at
large photon virtuality $Q$ and small hadronic mass $W$. Its amplitude
factorizes
in a $3 \pi$ generalized distribution amplitude and a hard subprocess $\gamma^*
\gamma \to q \bar q $ or $\gamma^* \gamma \to g g $. The interference of this
process with a Brehmsstrahlung process in $e e$ or  $e \gamma$ collisions
gives rise to interesting single spin asymmetries, particularly
 when $W$ is near the $\omega$ mass.

\vskip 1.5\baselineskip


\noindent
{\bf 1 Introduction.}
The process $\gamma^\ast \gamma \to \pi \bar \pi $ with a highly virtual
photon but small hadronic invariant mass $ W$ was recently studied in the
framework
of QCD factorization\cite{DGPT,DGP,fac}. It allows to study the pion pair
produced
in the isoscalar channel,  where the huge $\rho$-meson peak is absent. This
process is analogous to the single pion production, described by the
pion transition form-factor, being the long time object of QCD studies
\cite{BL,RR}.  In particular, the generalized
distribution amplitude (GDA), describing the non-perturbative stage
of this process, is a natural counterpart of the pion light cone
distribution amplitude, up to a notable difference. Namely, while the pion
distribution amplitude is a symmetric function of quark and antiquark
light cone momentum fractions, the relevant two pion
GDA is an antisymmetric one. This is a source  of the numerical
smallness when the convolution with hard scattering kernel
(also antisymmetric) is calculated. Namely, while integrated product
for the symmetric functions is small only at the edge of phase space,
when either quark or antiquark momentum fraction is small, for
antisymmetric functions it is also small in the region where their
momenta are approximately equal. More formally, this corresponds
to the contribution of higher Gegenbauer polynomials, similarly to the
case of Chernyak-Zhitnicki light-cone distribution\cite{CZ}.
As a result, the two-pion cross-section, while still experimentally
accessible, is suppressed with respect to the single-pion one by
an order of magnitude \cite{DGP}.

One may therefore ask whether it is possible to look for a GDA
which is still symmetric. The natural example is provided by
its generalization to the three meson case \footnote{This process was
comprehensively studied recently \cite{knecht} due to it sensibility
to the mechanism of chiral symmetry breaking.}.
This procedure, although straightforward,
possesses some interesting new features. The limiting case where one or two
of the  three mesons are soft may be described with the help of chiral
invariance.
The latter happens also to constrain the asymptotic
behaviour of some $3 \pi $ GDA with respect to the QCD evolution.
While the cross-section is basically of the order of the single pion
one, the observation of the extra pions allows to study
interesting angular and spin asymmetries.
In particular, interfering this process with the brehmsstrahlung of
a $\omega$ meson in $e e$ or $e \gamma$ collisions
  gives rise to a number of single spin asymmetries;
  some of them could be studied in future high statistics
  $e e$ experiments at medium energies.

\vskip\baselineskip
\noindent
{\bf 2 Kinematics.}

Let us consider the reaction $\gamma ^*(q) \gamma (q') \to \pi(p_1) \pi(p_2)
\pi(p_3)$. We define
$$
P= p_1 + p_2 +p_3,~~~ Q^2 = -q^2,~~~
W^2 = P^2
$$
and introduce the
lightlike vectors $v = (1,0,0,1) /\sqrt{2}$ and $v' = (1,0,0,-1)
/\sqrt{2}$ which respectively set a ``plus'' and a ``minus''
direction. We then have
\begin{eqnarray}
q   &=& {Q\over\sqrt{2}}\,(v-v')  \eqcm \hspace{4em}
q'   =  {Q^2+W^2\over\sqrt{2}\,Q}\,v'  \eqcm \nonumber \\
\PP &=& {Q\over\sqrt{2}}\,v + {W^2\over\sqrt{2}\,Q}\,v'  \eqpt
\end{eqnarray}

\noindent
In the kinematical region considered, $ W^2 << Q^2 $, $\PP$ lies mostly
along the  ``plus'' direction. The three  pion case may be considered in
complete analogy with the  two-pion one. Although the invariant
amplitude depends in general on the three additional subchannel
energies squared
$W_{ij}^2=(p_i+p_j)^2$, the situation is remarkably simplified when
one neglects the transverse momenta of the produced pions and
consider them to be almost parallel
so that one may neglect their "minus" components.
Consequently, the pion momenta $p_i$ are defined by their light-cone
fractions $\zeta_i$.
\begin{equation}
  \label{zeta}
p_i^+= \zeta_i P^+,
\end{equation}

\vskip\baselineskip
\noindent
{\bf 3 Expression for the amplitude.}

Following the  analogy with the two-pion case, the amplitude factorizes
as a convolution of a hard scattering kernel and a soft matrix element,

\begin{eqnarray}
\label{def}
i T^{\mu\nu} &=& - \int\! d^4 x\, e^{-i q\cdot x} \,
    \out\langle \pi \pi \pi |\,
    T J_{\mathrm{em}}^\mu(x) J_{\mathrm{em}}^\nu(0) \,| 0 \rangle\in
    \nonumber \\
&=& \int\! d z\, H^{\mu\nu}_{\alpha\beta}(z,q,q') \,
    S_{\alpha\beta}^{\phantom{\mu}}(z,v',p,p') \eqcm
\end{eqnarray}

Here the hard kernel is essentially the same as the one entering the
two-pion (and single pion) production,

\begin{eqnarray}
H_q^{\mu\nu} &=& {i e_q^2\over\sqrt{2}\, Q}\left\{
\left( g^{\mu \rho} v'^\nu + v'^\mu g^{\nu \rho}
       - g^{\mu\nu} v'^\rho \right) \gamma_\rho\,
     {2z-1\over z(1-z)} \right.  \nonumber \\
&& \left. \phantom{{i e_q^2\over\sqrt{2}\, Q}}
   - i \epsilon^{\mu\nu\rho\sigma}
   \gamma_\rho \gamma_5
   \left(  v_\sigma - {v'_\sigma\over z(1-z)} \right)\right\}
\end{eqnarray}

while
the large distance matrix element is

\begin{eqnarray}
\label{soft-S}
S_{\alpha\beta} &=& {P^+ \over2\pi} \int\! dx^-\,
  e^{-i z (P^+ x^-)} \,
\cdot   \nonumber \\
&&
   \out\langle \pi \pi \pi |\,
   \bar{\psi}_\alpha(x^- v')\psi_\beta(0) \,| 0 \rangle\in  \eqpt
\end{eqnarray}

Adding the third pion changes its  Dirac structure.
In fact, it is more
similar to the pion light-cone distribution than to the $ 2 \pi$ GDA,
because of the odd number of pions (pseudoscalar particles) involved.
At the leading twist level
there is only one leading  $ 3 \pi$GDA, which is completely analogous to
the pion light-cone  distribution.

\begin{equation}
S_{q, \alpha \beta}^{\phantom{\mu}}\,  \gamma^{+}_{\alpha \beta} \gamma_5
   =   \frac{i}{f_\pi}\dist_q (z, \zeta_1, \zeta_2,  \zeta_3~;
W_{12}^2,  W_{13}^2,  W_{23}^2,) \, P^+
\end{equation}
Here we have chosen the phase and scale of the matrix element
in accordance with a soft-pion theorem (see below), so that
the relations between various distributions will  contain
neither phase nor $f_\pi$.

The three light cone fractions are normalized by the condition
$\zeta_1 +
\zeta_2 +  \zeta_3 =1$, making only two of them independent,
while the squared total energy of the three pions is
$W^2=W_{12}^2+W_{13}^2+W_{23}^2 -3m_\pi^2$, where $m_\pi$ is the pion mass.

Conntrarily to the case of inclusive collinear
fragmentation, one is naturally interested in the transverse
components of the  produced pions momenta.
This change emerges
already for the two-pion GDA, where one might consider the relative
momentum of the two pions
$$
\Delta P=p_1-p_2,~~~~ \Delta P \cdot P=0~
$$
in complete analogy with the polarization vector of the
vector meson, should these two pions form a resonance. Consequently, one
may consider the transverse components of $\Delta P$ in an analogy
with the transverse polarization of a vector meson. Recall, that
the latter is
described   by the leading twist chiral-odd light-cone distribution
and by a chiral even term which is kinematically suppressed and
receives the contributions of twists 2 and 3, in complete analogy with
the structure function $g_2$ in DIS.

To avoid these complications, let us for the time being
neglect the transverse components
of pion momenta, this approximation being analogous to
considering the longitudinally polarized vector mesons only,
which is known to be the most easily described in the framework of QCD
factorization \cite{coll}.

In this approximation, the expression for the amplitude is
similar to that for the pion transition form-factor~:

\begin{eqnarray}
\label{final-tensor}
T^{\mu\nu} &=& {1\over 2 f_\pi}
\epsilon^{\mu\nu\rho\sigma}
v^\rho v'^\sigma \,
\sum_q e_q^2 \,
   \int_0^1 dz\, {1\over z(1-z)} \,
   \dist_q(z, \zeta, W^2)  \eqpt
\end{eqnarray}

\vskip\baselineskip
\noindent
{\bf 4 Three pion generalized distribution amplitudes.}

Many properties of the three-pion distribution amplitudes are common to the
two-pion
case. In particular, their evolution is governed by the same ERBL dynamics
\cite{ERBL},
as in the single pion and two pion cases, as the short distance
subprocess is the same in all the cases.

Analogously, charge conjugation invariance provides a symmetry
relation, which for the case of $\pi^0 \pi^+ \pi^- $ is
\begin{equation}
\label{symmetry}
\dist(z, \zeta_0, \zeta_+,\zeta_-) = \dist(1-z, \zeta_0,
\zeta_-, \zeta_+) \eqcm
\end{equation}
While for the case of $3\pi^0$ the function should be separately
symmetric in the interchange $z \to 1-z$ and with respect to the
interchange of any
two $\zeta_i$:

\begin{eqnarray}
\label{symmetry1}
\dist(z, \zeta_1, \zeta_2,\zeta_3) = \dist(1-z, \zeta_1,
\zeta_2, \zeta_3)=
\dist(z, \zeta_2, \zeta_1,\zeta_3 ) \nonumber \\
=\dist(z, \zeta_3, \zeta_2,\zeta_1)
=\dist(z, \zeta_1, \zeta_3,\zeta_2)
\eqcm.
\end{eqnarray}

As two-photon collisions  select just the symmetric part
of the amplitude, we will limit ourselves to its study in the charged pions
case.

One of the main properties of the three-pion amplitude in real
photons collisions, discussed in the seminal paper \cite{altz},
is the low energy theorem which states that the amplitude
vanishes in the  soft neutral pion limit.
The main reason of that property is the fact that the isovector axial
current, being the interpolating field for the $\pi^0$,
commutes with the electromagnetic current.
As the matrix element under investigation is also
represented as a matrix element of the product of
electromagnetic currents (\ref{def}), the theorem applies also to
the virtual photon case.

The theorem \cite{altz} is actually valid for any number of pions,
and, in particular, for the most simple two-pion case. Its validity
is then guaranteed by the low energy theorems for the two-pion GDA,
relating its value for $\zeta \to 0,1$ to the pion light-cone
distribution \cite{otb}. As the latter is symmetric in $z$,
it  may appear in the isovector
channel only, while the antisymmetric in $z$
isoscalar contribution and in particular
the whole $2 \pi^0$ amplitude is zero in that limit \cite{Pol}.
As a result the low energy theorem \cite{altz} is satisfied.

One should have a similar realization of
this theorem through the properties of the 3-pion GDA.
Putting the neutral meson
momentum to zero, one gets:

\begin{equation}
\dist_q (z, 0, \zeta_2,  \zeta_3,
0,  0,  W_{23}^2 )=0,
\end{equation}
where we labeled the $\pi_0$ meson first, while the two other mesons
can be either $\pi^+ \pi^- $ or  $\pi^0 \pi^0 $.

At the same time, putting both charged pion momenta to zero,
one should get the GDA in this special situation equal to the pion
distribution amplitude,

\begin{equation}
\label{2low}
\dist_q (z, \zeta_0, \zeta_2=\zeta_3=0~;
0,  0,  0)=\varphi_\pi(z),
\end{equation}
where $\varphi_\pi(z)$ is normalized to unity.
These restrictions are  especially useful when one takes into
account the polynomiality condition \cite{Pol,ji}, which reflects the fact
that,
after local expansion of (\ref{soft-S}),
the entire $\zeta$ dependence  comes from the contraction of space
separation vectors with momenta in the tensor decomposition of local
operators.

The application of this condition is more restrictive in the case
of the asymptotic  $z$ dependence, where only a small number of local
operators contribute.
In particular, assuming the asymptotic $z$ dependence $z(1-z)(1-2z)$
for the 2-pion isoscalar GDA results, by imposing the low energy theorem,
in a particularly simple $\zeta$ dependence

\begin{equation}
\dist_q (z,\zeta)=N \zeta(1-\zeta)z(1-z)(1-2z).
\end{equation}

For $\pi^0 \pi^+ \pi^-$     distribution the asymptotic $z$-dependence,
because of the different symmetry, is just
\begin{equation}
\label{symmetry2}
\dist_q (z, \zeta_0, \zeta_+,\zeta_-) = \frac{1}{6(1+a)}
z (1-z) (\zeta_0+a \zeta_0^2),
\end{equation}

where $a$ is an unknown parameter and the normalization has been fixed with
the help of (\ref{2low}).
This is the only quadratic polynomial compatible with the low energy
theorem for $\pi^0$. This means, in particular, that it is impossible
to have a non-zero asymptotic GDA for three neutral pions. The  $\pi^0
 \pi^0 \pi^0$ production in
$\gamma \gamma^*$ collisions is then logarithmically suppressed at
large $Q^2$ with respect to $\pi^+ \pi^- \pi^0$ production.



\vskip\baselineskip
\noindent
{\bf 5 Cross-sections and Single Spin Asymmetries.}

The similarity between the three--pion GDA and the single-pion DA
is leading to a simple relation between the single-pion and three-pion
cross-sections. Namely, the ratio of matrix elements is
proportional to the ratio of distribution amplitudes, so that~:

\begin{eqnarray}
\label{comp}
T^{\mu\nu}_{\pi} &=& {f_\pi\over 2}
\epsilon^{\mu\nu\rho\sigma}
v^\rho v'^\sigma \,
\sum_q e_q^2 \,
   \int_0^1 dz\, {1\over z(1-z)} \,
   \phi (z)  \eqpt     \nonumber \\
T^{\mu\nu}_{3\pi} &=& {1\over 2 f_\pi}
\epsilon^{\mu\nu\rho\sigma}
v^\rho v'^\sigma \,
\sum_q e_q^2 \,
   \int_0^1 dz\, {1\over z(1-z)} \,
   \dist_q(z, \zeta, W^2)  \eqpt
\end{eqnarray}

It then follows that

\begin{equation}
\label{rat}
T^{\mu\nu}_{3\pi} = \frac{N}{f^2_\pi} T^{\mu\nu}_{\pi}
\end{equation}
with
\begin{equation}
N= \frac{\int_0^1 dz\, {1\over z(1-z)} \,
   \dist_q(z, \zeta, W^2)} {\int_0^1 dz\, {1\over z(1-z)} \,
   \phi (z)}
\end{equation}

Assuming that both $\phi(z)$ and
$\dist_q(z, \zeta, W^2)$ are close to their asymptotic forms, one may
conclude  in the spirit of low-energy theorem that N is not far from 1.

The extra mass dimension in the three-pion distribution should, of
course, be compensated by the phase space of the two additional pions,
so that

\begin{eqnarray}
\frac{d\sigma_{3 \pi}}{d\sigma_\pi} \sim N^2
\frac{W_{max}^4 }{16 \pi^2 f^4_\pi}
\label{ratsi}
\end{eqnarray}
where $W_{max}$ is an upper value of hadronic invariant mass.
In the case  where it is of order of 1 GeV, one gets a
ratio of order $0.1-1$.

Note that such a similarity between production of three and
one pion is entirely due to the participation of a virtual photon.
In the real photon case there is a strong destructive interference between
invariant amplitudes, reducing the cross-sections by two orders of
magnitude \cite{knecht}. In our case, only one of these amplitudes,
proportional to the single pion one, survives in the limit of the
asymptotically large $Q^2$. Consequently, in the region
of moderate $Q^2 \sim 1-2 GeV^2$ one should expect that subleading
contributions should provide negative corrections to the cross-section,
decreasing the above estimate.

The existence of three hadrons in the final state
increases the number of observables, sensitive to GDA.
We are suggesting here to use for such purpose the 3 pion
{\it handedness}. The latter is known to be the T-odd observable,
introduced originally to describe the consequense of fragmenting
polarized partons in the distribution of produced spinless particles
\cite{EMT}.

\begin{eqnarray}
H=\frac{N_+- N_-}{N_+ + N_-}
\label{hand}
\end{eqnarray}
where $N_\pm$ is the number of pion triplets with the mixed product
$(\vec p_1 \times \vec p_2) \cdot p_3$ of their momenta being
positive or negative, respectively.

At the same time, it may also be used to describe polarization effects
in exclusive reactions. In particular, the handedness in $e^+ e^-$
annihilation to four pions was studied in the framework of
chiral dynamics \cite{bkst}. The manifestation of the
handedness in the low energy three pion
production is at the same time suppressed. The reason relies on the
specificity of handedness, which is an example of Single Spin Asymmetry
(SSA).
For such asymmetries
the interference of various amplitudes is required to have a T-odd
effect, while the dominant amplitude through $\omega$ meson resonance
has a unique tensor structure. The discussed 3-pion GDA is therefore
providing a necessary ingredient to observe the handedness.

A similar asymmetry may emerge in the case of the two-pion production as
well,  where it should be due to the interference with a brehmsstrahlung
background, which is in that case dominated by the $\rho$ meson \cite{DGP}.
As soon as the dominant contribution to the two-pion GDA is provided by
the $\sigma$ meson \cite{DGPT}, one may speak in this case
of $\rho-\sigma$ interference, making such an effect an exclusive
analog of the T-odd fragmentation function, introduced by Jaffe, Jin
and Tang \cite{JT}.

The kinematical structure of the two-pion SSA is actually unique.
There are only four independent vectors
in the case when  one of the colliding leptons is longitudinally
polarized, which uniquely define the pseudosclar, required to
get a nonzero SSA.
These are the momenta of photons $q$ and $q'$, the relative pion
momentum
$\Delta P$, and the final momentum of the hard scattered lepton (while the
momenta
of the real photon and its parent electron are almost collinear).

The kinematical structure of the three-pion
SSA is richer. While the momenta of all three pions should
naturally enter the  pseudoscalar implied by the structure
of the interfering  $\omega$ meson dominated amplitude,
the choice of the fourth momentum
is left free. It can be the momentum of any of the colliding leptons.
This may be considered as two ways of defining handedness,
with respect to two  rest frames, differing  by a Lorentz boost
along the $z$ direction.

The participation of the $\omega$ meson allows to make important
simplification in the consideration of the three-pion GDA,
as anticipated in Section 2. Since the  $\omega$ meson
form-factor already includes the transverse components of pion
momenta, it is possible, in the leading approximation,
to consider the purely  longitudinal GDA.
The spin-dependent part of the cross-section is therefore
proportional to~:

\begin{eqnarray}
\label{dels}
\Delta \sigma  \sim  Im(F_\omega \Phi^*_q)\{
\epsilon^{k, p1, p2, p3} [-\lambda_1 (x_B^3+3 x_B^2+2 x_B)
+\lambda_2(2+ x_B^3+3 x_B^2+4 x_B)] \nonumber \\
+\epsilon^{p, p1, p2, p3} [-\lambda_1 (x_B^3+2 x_B^2)
+\lambda_2(x_B^3+2 x_B^2+2 x_B)]\}
\end{eqnarray}
\noindent
where the usual notation
$\epsilon^{a,b,c,d} =
\epsilon^{\alpha \beta \gamma \delta} a_\alpha b_\beta c_\gamma d_\delta$
has been used. Here
$x_B=Q^2/2p\cdot k$,  $k,p$ being the momenta of colliding
(real) photon and electron, while
$\lambda_{1,2}$ are the longitudinal  polarizations of electron and
circular polarization  of the real photon, respectively, so that two
single asymmetries are arizing from this expression.

One may compare these asymmetries with SSA in Deeply Virtual
Compton Scattering on a pion \cite{ji,G}, related to them by crossing.
While its kinematical structure
is quite similar, the dynamical origin of the phase shift is different.

While in DVCS the imaginary part  is produced by the quark propagator in
the hard scattering subprocess going on-shell, the phases in the $\gamma^*
\gamma $ process are due to the phase
shifts between time-like meson form factors and
the imaginary parts of GDA, which naturally  emerge at the
non-perturbative level \cite{DGPT}. It happens that only the
latter is essential for the following reason.

To observe a sizable asymmetry one should concentrate on
the kinematical
region where  the interference term
$Im(F_\omega \Phi^*_q)$ is  not significantly
suppressed
 with respect
to non-interference terms, and, in particular, with respect to the squared
Brehmsstrahlung contribution $|F_\omega|^2$.
This means, that the invariant mass of the three pions should not be too
close to the mass of the $\omega$ meson.
Since the imaginary part of the $\omega$ meson form-factor in that
region  is fastly decreasing, it seems reasonable to keep only its real
part  in the interference term. This, in turn,  means that
only the imaginary part of the
GDA contributes to the SSA. The situation is somehow similar to
what happens in perturbative calculation, where the Born term is purely
real and only the imaginary part of one-loop amplitude produces SSA.

To exploit this idea further, one may define the SSA in such a way,
that only the interference term contributes to the spin averaged
cross-section as well. This may be achieved by considering the
charge asymmetry \cite{charge} and the following definition of the SSA~:
\begin{equation}
A=\frac{d \sigma_{e+}-d \sigma_{e-}-d \sigma_{\bar e +}+
d \sigma_{\bar e -}}
{d \sigma_{e+}+d \sigma_{e-}-d \sigma_{\bar e +}-
d \sigma_{\bar e -}},
\end{equation}
where, say $d \sigma_{e+}$ corresponds to the differential
cross-section initiated by the electrons with positive helicity, and
so on. The $\omega$ form-factor in the numerator and denominator
cancels, so that~:

\begin{eqnarray}
\label{Afin}
A_e =- Q^2 \frac{Im(\Phi_q)}{Re(\Phi_q)} \cdot
\frac{\epsilon^{k, p1, p2, p3} (x_B^2+3 x_B+2)
+\epsilon^{p, p1, p2, p3}(x_B^2+2 x_B)}{2
\epsilon^{\mu, p1, p2, p3} \epsilon^{\mu, k, p, q}
(2+x_B^2-2 x_B)}\nonumber \\
A_\gamma = Q^2 \frac{Im(\Phi_q)}
{Re(\Phi_q)}
\cdot
\frac{[\epsilon^{k, p1, p2, p3}(2+ x_B^3+3 x_B^2+4 x_B)
+\epsilon^{p, p1, p2, p3} (x_B^3+2 x_B^2+2 x_B)]}
{2x_B
\epsilon^{\mu, p1, p2, p3} \epsilon^{\mu, k, p, q}
(2+x_B^2-2 x_B)},
\nonumber
\end{eqnarray}
where $A_{e,\gamma}$ are the asymmetries for polarized lepton and
photon, respectively.

One does not expect any specific parametric
suppression of the imaginary part of three-pion GDA with respect to
the corresponding one for
two pions, which is well-known from phase shift analysis
\cite{Pol}.

Both two-pion and three-pion asymmetries may be measured in the
forthcoming high statistics two-photon experiments
at BABAR and BELLE. One should note, that two-pion asymmetry is completely
analogous to the azimuthal asymmetry recently discovered at
HERMES\cite{Hermes} and
may be considered as its exclusive analog.

\vskip\baselineskip
\noindent
{\bf 7 Conclusion.}
We generalized to the three pion case the notion of generalized distribution
amplitude. This extension of the QCD factorization properties to the
reaction $e e
\to e e \pi \pi \pi $ is a promising way to extract interesting new
information about
the hadronization of a $q \bar q$ pair.

A suitable tool to study this GDA is the single spin asymmetry,
which is linear in it and contains only its imaginary part.

\vskip\baselineskip
\noindent
{\bf Acknowledgments.}
We acknowledge useful conversations with M.~Diehl, A.V.~Efremov, 
N.A. Kivel, M.~Knecht and M.V.~Polyakov.

\end{document}